\documentclass{ws-ijmpb}
\usepackage{color}
\begin{document}
\clearpage
%
\catchline{}{}{}{}{}
%
\title
{Bose-Einstein condensate in a quartic potential: Static and Dynamic properties}
\author{G. K. Chaudhary, Amit K Chattopadhyay, and R. Ramakumar}
\address{Department of Physics and Astrophysics, University of Delhi,
Delhi-110007, Delhi, India}
\maketitle
\begin{history}
\received{9 May 2009}
\end{history}
\begin{abstract}
In this paper, we present a theoretical study of a Bose-Einstein condensate
of interacting bosons in a quartic trap in one, two, and three dimensions. Using
Thomas-Fermi approximation, suitably complemented by numerical
solutions of the Gross-Pitaevskii equation, we study the ground sate
condensate density profiles, the chemical potential, the effects
of cross-terms in the quartic potential, temporal evolution of various
energy components of the condensate, and width oscillations of the condensate.
Results obtained are compared with corresponding results
for a bose condensate in a harmonic confinement.
\end{abstract}
\section{Introduction}
\label{sec1}
Since the first experimental observation of Bose-Einstein (BE) 
condensation\cite{cornell1,davis,bradley}
in bose atom clouds in a harmonic trap, both experimental and theoretical 
studies of this phenomenon and various properties of the 
condensate has grown up 
to a fast expanding field of 
research \cite{dalfovo,leggett,anglin,pethick,bloch1,bloch2,yukalov}.
While studies of Bose condensation in a harmonic potential 
is of great interest, investigations of the BE condensation of lattice-bosons
and free-bosons in {\em anharmonic} potentials, in particular 
in a quartic potential, have received considerable attention 
in recent times\cite{gygi,ram,lundh,pnma,gautam} (see also the note 
in Ref. \cite{mixedtraps}). 
All the previous works\cite{gygi,ram,lundh,pnma} , except 
two\cite{bagnato,gautam}, 
deals with {\em lattice-bosons} in anharmonic confinements.  
The previous works on free-bosons in power law traps
and quartic traps in three and lower dimensions\cite{bagnato,gautam}
dealt with the calculations of Bose condensation
temperature for {\em non-interacting} bosons in those environments.
In this paper, we consider {\em interacting} bosons in a quartic potential.
The study of {\em interacting} bosons in a quartic potential 
may have important implications for fields ranging from chaotic 
dynamics to high energy physics\cite{gautam}.
Further, it would be desirable to have minimum inhomogeneity in the density
profiles for clear observation of several phases and transitions 
possible for interacting boson, fermion, and boson-fermion models\cite{bloch2,yukalov}.
While interactions in a many-boson system are expected to
broaden the real space distribution of bosons in the confining region,
it is not clear how far, from the center, a homogeneous distribution
of bosons will exist. We will show, among other results, that the
interaction leads to a more or less homogeneous boson distribution
in a substantial region (compared to the non interacting bosons case)
in and around the trap center. It would also be of contemporary interest
to investigate the dynamics of the condensate in the specific quartic potential
environment and compare it with the usually employed quadratic
environment. It may be expected that the experiments would first 
involve investigations of interacting bosons in a quartic potential
before an additional lattice potential is applied.
Furthermore, experimental studies of interacting bosons in the simplest
anharmonic environment may be of considerable interest in its own right.
Within this general setting, it would be useful to make a detailed theoretical
study of {\em interacting} bosons in a quartic potential.
It is the purpose of this paper to present some results of 
such an investigation using a combination of analytical and numerical methods.
\par
This paper is organized in the following three sections.
In Sec {\ref{sec2}}, the condensate properties are analyzed 
through a study of 
the Gross-Pitaevskii equation\cite{gross,pitaevskii} employing
the Thomas-Fermi approximation\cite{thomas,fermi}.
In Sec {\ref{sec3}} we present a comparative study of real space
distributions of bosons in condensates of interacting bosons in quartic and
harmonic confinements, by numerically
solving the Gross-Pitaevskii equation. The time evolution 
of some properties (various components of energy and width oscillations
of the condensate) of the condensate
is given in Sec {\ref{sec4}}, and the conclusions are presented in Sec 
{\ref{sec5}}.
\section{BEC in a quartic trap: Thomas-Fermi approximation}
\label{sec2}
The ground state properties of a Bose-Einstein condensate (BEC)
is well described by a macroscopic wave-function $\psi({\bf r},t)$, 
the time evolution of which is governed by the 
Gross-Pitaeveskii equation \cite{gross,pitaevskii} (GPE).
Incorporating the trap potential as well as a repulsive interaction 
between the atoms 
forming the condensate, the GPE in three dimensions (3d) is given by
\begin{eqnarray}
 i \hbar\frac{\partial\psi({\bf r},t)}{\partial t}&=&-\frac{\hbar^2}{2m}\nabla^2\psi({\bf r},t)+
v({\bf r})\psi({\bf r},t) 
+ NU_{0}\vert\psi({\bf r},t)\vert^2\psi({\bf r},t)\,,
\label{NLS_eqn}
\end {eqnarray}
where ${\bf r} \equiv (x,y,z)^T$, $m$ is the atomic mass, $v({\bf r})=\alpha {({\bf r}.{\bf r})}^2$  
is the quartic trapping potential, $N$ is the total number of atoms forming the
condensate, $U_{0}(=4\pi\hbar^2a/m)$ is the strength of 
interaction between atoms in the condensed state, and $a$ ($> 0$) is 
the s-wave scattering length. 
\par
Before applying the Thomas-Fermi approximation, we non-dimensionalize 
Eq. (\ref{NLS_eqn}) through a set of linear 
transformations: $\tilde{t}=\omega t,\:\:\tilde{{\bf r}}={\bf r}/l,\:\:
\widetilde{\psi}({\bf r})=l^{3/2}\psi({\bf r})$. After dropping the wiggles on the symbols, 
we obtain  
\begin{eqnarray}
 i \frac{\partial\psi({\bf r},t)}{\partial t}&=&-\frac{1}{2}\nabla^2\psi({\bf r},t)+ \gamma\:
 {({\bf r}.{\bf r})}^2\psi({\bf r},t) 
+\lambda\: N\: \vert\psi({\bf r},t)\vert^2\psi({\bf r},t)\,,
\label{NLS_non_dim_eqn}
\end {eqnarray}
where
\begin{equation}
l=\sqrt{\frac{\hbar}{m \omega}},\,\, \lambda = \frac{4\pi a}{l},\,\,\:\:\gamma=\frac{\alpha \hbar}{m^2 \omega^3}.
\end{equation} 
In the absence of the non-linear interaction term, the GPE is 
reduced to a Schr\"odinger equation (with a quartic  
potential) whose approximate eigenvalues can be obtained through perturbation theory or
WKB approximation \cite{mathews}. Expressing $\alpha$ in units of 
$m^2\omega^3/\hbar$, 
we perform dimensional reductions \cite{williams} to obtain corresponding 
GPEs in lower dimensions.
In general, the GPEs for d = 1, 2, and 3 can be summarized in a single equation as
\begin{eqnarray}
 i \frac{\partial\psi({\bf r},t)}{\partial t}&=&-\frac{1}{2}\nabla^2\psi({\bf r},t)+
 {({\bf r}.{\bf r})}^2\psi({\bf r},t) 
+k_{d} \vert\psi({\bf r},t)\vert^2\psi({\bf r},t)\,,
\label{NLS_non_dim_eqn}
\end {eqnarray}
where
\begin{equation}
k_{d}=\alpha_{d}\lambda_{d} N,\; \:\: \lambda_{d} = \frac{4\pi a}{l^{d-2}}.
\label{constants}
\end{equation} 
Here $\alpha_{2}$ and $\alpha_{1}$ are scaling factors introduced 
to write GPE in 2d and 1d, respectively.
In order to find a stationary solution of 
Eq. (\ref{NLS_non_dim_eqn}), we do a separation of variables
$\psi({\bf r},t)=\psi({\bf r}) \times exp[-i (\mu/(\hbar\omega))t]$,
where $\mu$ is the chemical potential. Starting 
from Eq. (\ref{NLS_non_dim_eqn}), we obtain
\begin{equation}
-\frac{1}{2}\nabla^2\psi(r)+r^{4}\psi(r)+k_{d}\vert\psi(r)\vert^2\psi(r)=\frac
{\mu}{\hbar\omega}\psi(r)\,.
\label{stationary_state_eqn}
\end {equation}  
In the preceding equation $r^{4} \equiv x^{4}+y^{4}+z^{4}$. 
For simplicity we have neglected the cross terms from the 
potential (see, however, Sec. III-C where these terms are shown to have
only minor effects on the properties we study). 
To analytically solve for the ground state of the stationary state 
GPE, as given in Eq. (\ref{stationary_state_eqn}), we make use 
of the Thomas-Fermi approximation \cite{dalfovo,thomas,fermi,baym}. For 
sufficiently large number of atoms, the kinetic energy is very small compared to
the potential and the interaction energies. So, a good approximation to the 
ground state is obtained by solving the GPE without the kinetic 
energy term. On using this approximation, the GPE becomes 
\begin{equation}
r^4\:\psi(r)+k_{d}\vert\psi(r)\vert^2\psi(r)=\frac{\mu}{\hbar\omega}
\psi(r),
\end{equation}
The relevant solutions of the preceding cubic equation are
\begin{equation}
\psi(r) = \pm \sqrt{\frac{1}{k_{d}}\left( \frac{\mu}{\hbar \omega}-r^4 
\right)}.
\label{cubic_eqn}
\end{equation}
Since $\psi(r)$ vanishes for $\frac{\mu}{\hbar\omega} \leq r^{4}$, the 
Thomas-Fermi radius $r_{\mathrm{TF}}$ is obtained as
\begin{equation}
r_{\mathrm{TF}}=\left(\frac{\mu}{\hbar\omega}\right)^{1/4}
\end{equation}
in which the chemical potential ${\mu}_{\mathrm{TF}}$ is be determined
from the normalization condition 
\begin{equation}
c_{d}\int_{0}^{r_{\mathrm{TF}}}\vert\psi(r)\vert^2\:r^{d-1}dr=1\,,
\label{normalization}
\end{equation}
where $c_{d}$ = $2$, $2\pi$, and $4\pi$ for $d$ = 1, 2, and 3, respectively. 
Then, the chemical potential is found to be
\begin{equation}
{\mu}^{d}_{\mathrm{TF}}=\hbar\omega\left[\frac{d(d+4)}{4 c_{d}} k_{d} \right]^\frac{4}{d+4}.
\end {equation}
So, for the 3d case, the chemical potential varies as  
$N^{\frac{4}{7}}$ for the quartic trap. In comparison, we note that
it varies as $N^{\frac{2}{5}}$ for bosons in a harmonic trap\cite{williams,bao}
(for which $v({\bf r}) \propto {\bf r.r}$).
It may be noted that the $N$ dependence of the chemical
potential becomes more dominant as the dimensionality {\em decreases}.
\par
The scaling factors ($\alpha_{2}$ and $\alpha_{1}$) were introduced in 2d and 
1d GPE to obtain GPEs in 2d and 1d. They also ensure that the size of the 
condensate remains the same in all the three dimensions. 
To find $\alpha_{2}$ and $\alpha_{1}$, we calculate 
$\mu_{TF}$ in 2d and 1d and equate it to the chemical potential 
obtained in 3d\cite{williams}. Using this procedure, we obtain
\begin{equation}
\alpha_{3}=1\,,
\end{equation}
\begin{equation}
\alpha_{2}=\frac{1}{6}(\frac{21}{4})^\frac{6}{7}N^\frac{-1}{7}(a{l}^6)^{-\frac{1}{7}}\,,
\end{equation}
and
\begin{equation}
\alpha_{1}=\frac{2}{5 \pi}(\frac{21}{4})^\frac{5}{7} N^\frac{-2}{7}(a^2{l}^{12})^{-\frac{1}{7}}.
\end{equation}
\section{Spatial distribution of bosons in the condensate}
\label{sec3}
\subsection{Condensate density profiles in 1d}
In this section, we present results on the
spatial distribution of bosons in the condensate in 1d by numerically solving
the GPE. We solve the GPE by using
the finite difference Crank-Nicholson (FDCN) method\cite{williams,adhikari}.
In this method for the GPE given by 
\begin{equation}
i \frac{\partial\psi(x,t)}{\partial t}=H\psi(x,t)\,,
\label{gpe CN}
\end {equation}
the solution is advanced in small time steps $\delta t$  as given below
\begin{equation}
\psi(x,t+\delta t)=\frac{1-i\frac{\delta t}{2}H}{1+i\frac{\delta t}{2}H}\psi(x,t).
\label{CN1d}
\end {equation}
The ground state solution is found by the imaginary time propagation 
method\cite{williams}.
In the imaginary time propagation method, we replace $\delta t$  by $-i\delta t$
and propagate a initial trial wave function using the above mentioned 
propagation scheme. After suitable normalization
at each time step the wave-function converges to the ground state.
We have taken the space step as $\delta x=0.1$ and 
the time step as $\delta t=0.001$.
For illustrative purpose, we have used a set of parameters for ${}^{87}\textrm{Rb}$:
$m\,=\,1.44\times10^{-25}\,Kg$, $a\,=\,5.1\times10^{-9}\, m$,
$\nu\,(=\omega/2\pi)\,=\,24\,Hz$.
\par
In Fig. 1, we show the  condensate density profiles for systems with 
$10$ to $10^6$ bosons in a quartic trap. Further, this figure also contains
results of the TF approximation for the same numbers of bosons in a quartic
trap. Furthermore, we have also shown the results for bosons in a harmonic
potential. We find that the TF approximation results are in quantitative agreement
with the numerical solutions of GPE when the number of bosons is large. 
On increasing the number of atoms, the width of the ground state increases, 
which is similar to the case of bosons in a harmonic potential\cite{williams,bao}, 
except that the spread of condensate is larger for a harmonic potential. We also
notice that the ground state density profiles for the quartic trap becomes relatively
flat with increasing number of bosons in the quartic trap. 
Next,  we fix the number of atoms to $10^4$ in a quartic trap and investigate the boson
distribution in the ground state for different interaction strengths.
The results are displayed in the left panel of Fig. 2. We find that the ground
state boson distribution is  more spread out with increasing
interaction strength. It may be argued that the redistribution of bosons
from the center to other parts of the condensate is due to
the peak in the non-interacting distribution in the central region
since the $U=0$ wave-function is peaked at the center and is symmetric
around the center. But, for the
interacting bosons case, the interaction term is to be calculated
in a self-consistent manner including the effect of the interaction.
Hence, it is not obvious that interacting bosons distribution
will have the profile shown in Fig. 2. It is the result of a
combined effect of interaction and self-consistent changes
in the wave-function. Thus, we see that the interacting bosons,
compared to the non-interacting bosons,
have a density profile which is more or less uniform in the
central region in the case of a quartic confining potential.
In Fig. 2, we have also exhibited the ground state 
chemical potential for different 
values of the interaction parameter $k_{1}$. The chemical potential
for a condensate in 
a quartic trap is larger than that in harmonic trap, the 
difference becoming more significant for higher values of $k_{1}$.
\subsection{Condensate density profiles in 2d and 3d}
We have solved the GPE in higher dimensions by using the split-step 
FDCN method\cite{adhikari1}.
In this method, in 3d, the Hamiltonian $H$ is split into different 
non-derivative and
derivative  parts $H_{1},\,H_{2}$, $H_{3}$, and $H_4$, where
\begin{equation}
H_{1}=v({\bf r})+k_{3}\vert\psi({\bf r},t)\vert^2,\!\!\!\!\!\!\!\!\!\!
\label{non derivarive}
\end {equation}
\begin{equation}
H_{2}=-\frac{1}{2}\frac{\partial^{2}}{\partial x^{2}},\,\,
H_{3}=-\frac{1}{2}\frac{\partial^{2}}{\partial y^{2}},\,\,
H_{4}=-\frac{1}{2}\frac{\partial^{2}}{\partial z^{2}}.
\end {equation}
The time evolution is performed in the following steps. Let $\psi^{n}$
be the wave-function at time $t_{n}$. 
This wave-function is advanced first over a time step $\delta t$ at $t_{n}$ to produce 
an intermediate solution $\psi^{n+1/4}$ from  $\psi^{n}$  via 
\begin{equation}
\psi^{n+1/4}=e^{-i\delta tH_{1}}\psi^{n}.
\end{equation}
Next the time propagation is performed via the following semi-implicit Crank-
Nicholson scheme:
\begin{equation}
\psi^{n+2/4}=\frac{1-i\frac{\delta t}{2}H_{2}}{1+i\frac{\delta t}{2}H_{2}}
\psi^{n+1/4},
\end {equation}
\begin{equation}
\psi^{n+3/4}=\frac{1-i\frac{\delta t}{2}H_{3}}{1+i\frac{\delta t}{2}H_{3}}
\psi^{n+2/4},
\end {equation}
and
\begin{equation}
\psi^{n+1}=\frac{1-i\frac{\delta t}{2}H_{4}}{1+i\frac{\delta t}{2}H_{4}}
\psi^{n+3/4}.
\label{final time}
\end {equation}
An analogous set of equations hold in 2d. In the calculations,
we have used space steps of 0.1 (for $\delta x$ and $\delta y$
in 2d and $\delta x$, $\delta y$, and $\delta z$ in 3d) and the time 
step $\delta t=0.001$.
To calculate the ground state, again we use imaginary time propagation method
by replacing $\delta t$ with $-i\delta t$.
We start with an initial trial wave-function which is propagated via 
the above mentioned scheme. After suitable normalization,
at each time step, the wave-function converges to the ground state.
\par
In Fig. 3, the ground state density profiles of the condensate are 
shown for different numbers of atoms ranging from $10^3$ to $10^6$.
On increasing the number of atoms, the width of the ground state increases, 
which is similar to that in 1d. 
On comparing the ground states for harmonic and quartic traps as shown 
in Fig. 4, we found that the ground state density profiles are flat
in the central region for the quartic trap compared to the harmonic trap. 
Similar results are obtained for the 3d case as well.
Clearly, the quartic confinement is more advantageous to the quadratic one 
if one is interested in investigations of the properties of a bose 
condensate with a minimal influence of the confining potential. We may note 
here that, in an earlier
work\cite{ram}, we have shown that the finite temperature 
properties of {\em lattice}
bosons are closer to the pure lattice case if one uses quartic rather than
a quadratic confining potential.
The effect of changing interaction strength on the ground state density profiles
in a quartic trap are displayed in Figs. 5 and 6 for 2d and 3d, respectively.
It is clear that the bosons gain energy by spreading into larger area in 
the traps with increasing strength of the interaction.  
In Fig. 7, the ground state chemical potential is plotted for different 
values of the interaction parameter $k_{d}$ for $d$ = 2 and 3.
The chemical potentials of the
condensate in a quartic traps are higher than those in harmonic traps, the
difference becoming more significant at higher values of $k_{d}$.
\subsection{Effects of cross terms}
In the preceding section, we have presented results of our calculations
neglecting the cross terms in the full quartic potential 
($V({\bf r})= \alpha {({\bf r}.{\bf r})}^2$). In this section, we investigate
the effects of the previously neglected cross terms.
The contribution of the cross terms is in cross-coupling potentials 
across different dimensions. 
One may expect that the inclusion of these cross-terms may lead to
complex inhomogeneous density profiles.
But, on performing the calculations for the quartic trap,
we find that the effects of cross-terms are rather minor. They
lead to small increases
in the chemical potential and the number of bosons in the central
region of the trap as shown in Fig. 8 for 2d. We do not find
any significant inhomogeneities in the ground state condensate
profiles generated by the cross-terms. 
We have observed (figures not shown)
similar effects in 3d as well. It may be noted here that
cross terms lead to a $10\% $ increase of the bose condensation 
temperature ($T_c$) in 2d, and a $20\%$ increase in 3d\cite{gautam}.
\section{Time evolution of the condensate}
\label{sec4}
In this section, we study the time evolution of the condensate after
removal of the trapping potential and width oscillations of the condensate.
To theoretically study the expansion of condensate, we time evolve the GPE
after removing the potential term, taking the stationary ground state 
as the initial state function at $t=0$. To time evolve the GPE we use 
the FDCN scheme given by Eqs. \ref{gpe CN}-\ref{CN1d} in 1d and split-step FDCN 
scheme as given in Eqs. \ref{non derivarive}-\ref{final time} 
in higher dimensions. Here the propagation is done in real 
time with time-step $\delta t=0.0001$ and space steps 
of $0.1$ (for $\delta x$, $\delta y$, and $\delta z$). The different 
energy components we calculate are given by\cite{dalfovo,holland}:
\begin{eqnarray}
E_{\mathrm{kin}}(t)&=&-\frac{1}{2}\int\psi^*(r,t)\nabla^2\psi(r,t)dr,\nonumber\\
E_{\mathrm{pot}}(t)&=&\int\psi^*(r,t)v(r)\psi(r,t)dr,\nonumber\\
E_{\mathrm{int}}(t)&=&\frac{1}{2}\int\psi^*(r,t)k_{d}\vert\psi(r,t)\vert^2\psi(r,t)dr.
\end {eqnarray}
And, the the chemical potential is given by\cite{dalfovo,holland}:
\begin{eqnarray}
\mu(t)=\int\left(\frac{1}{2}\vert\nabla\psi(r,t)\vert^2+v(r)\vert\psi(r,t)\vert^2+k_{d}\vert\psi(r,t)\vert^4\right)dr
\end {eqnarray}
\par
The variation of various energy components and the chemical potential, for
bosons in quartic and harmonic traps, with time
is shown in Fig. 9 for 1d and 2d. Similar results (not shown) are 
observed in 3d.
As shown in these figures, the chemical potential and the 
interaction energy decreases on
time evolution, but there is an increase in kinetic energy term. 
Most  of the energy {\em before} the expansion ($t < 0$) is 
contained in the mean-field and the 
potential energy of the bosons. 
At $t=0$, the confining potential is removed and the corresponding energy 
term from then on is zero. During the expansion, the chemical potential 
decreases due to decrease in the mean field. Further, the increase in kinetic 
energy during the expansion is due to transfer from the mean-field. We have
also shown in Fig. 9, the release energy ($E_{\mathrm{rel}}=E_{\mathrm{kin}}+
E_{\mathrm{int}}$) which is a a time independent quantity.  
\par
The time evolution of the width of the condensate is displayed 
in Fig. 10 for 1d, 2d, and 3d. 
The condensate widths along three axes x, y, and z are defined as
$\Delta x = \sqrt{<(x-<x>)^2>}$, $\Delta y = \sqrt{<(y-<y>)^2>}$,
and  $\Delta z = \sqrt{<(z-<z>)^2>}$,
where  brackets $<>$ represent the expectation value at a particular instant
during the time evolution. We assume that at $t=0$, the condensate 
is in a non-interacting ground state. 
Once the interaction is switched on at $t=0$, it  
produces oscillations in the condensate.
The Fig. 10 show that the frequencies of oscillations are greater 
in the quartic 
potential case than in the harmonic case\cite{jak}. For isotropic quartic 
potential, amplitude and frequency of oscillations in x,y, and z directions 
are the same.
\section{Conclusions}
\label{sec5}
In this paper, we studied the static and dynamic properties of 
a BE condensate in a quartic trap in one, two, and three dimensions. 
The analysis was done using both analytical and complementary 
numerical techniques.
The analytical solutions were obtained using the Thomas-Fermi 
approximation. The solutions obtained were then compared with 
the results obtained by numerically solving the GPE.
We compared the ground state density profiles of the condensate in a
quartic trap for different strengths of interaction by changing the scattering 
length. It was found that the Thomas-Fermi approximation results were
in remarkably good agreement with the results obtained from the GPE for
a large number of atoms in the condensate in a quartic confining potential.
The results were also compared with corresponding results for
a bose condensate in a harmonic confining potential. 
We find that the condensate density profiles are relatively flat 
in the central region of the quartic potential compared to the 
case of a condensate in a harmonic potential. An implication of 
this result is that for investigating properties
of a bose condensate with minimal influence of the overall confining potential,
the quartic potential is more preferable compared to a quadratic potential.
We also find that increasing the number of atoms in the condensate or
increasing the interactions spreads the condensate in the confining region.
Our calculations of the chemical potential in the ground state show that
it is larger for a condensate in a quartic trap compared to that in
a harmonic trap. The preceding conclusions were arrived at without 
including the cross-terms in a general quartic potential. On including the 
cross-terms, we find that their
effects lead to increases in the chemical potentials and condensate
density in the central region of the trap. These changes were found to be
rather minor. From our results for the time evolution of the various energy 
components of an expanding condensate (after removing the confining
potentials), we find that the chemical potential and the potential energy 
decrease during the expansion while the kinetic energy increases.
It was also noticed that changes in the case of quartic potential 
are larger than the case of a harmonic potential. 
The condensate width oscillations produced by
switching on of the interaction, starting with a condensate of non-interacting
bosons, were also studied. It was found that the width oscillations of a
condensate in a quartic trap have higher frequency than that in a harmonic trap.
\newpage
\section{Acknowledgment}
\label{sec6}
Gopesh Kumar Chaudhary thanks UGC, Government of India, for  
financial support during this work. 

\begin{figure}
\resizebox*{4.0in}{4.0in}{\rotatebox{270}{\includegraphics{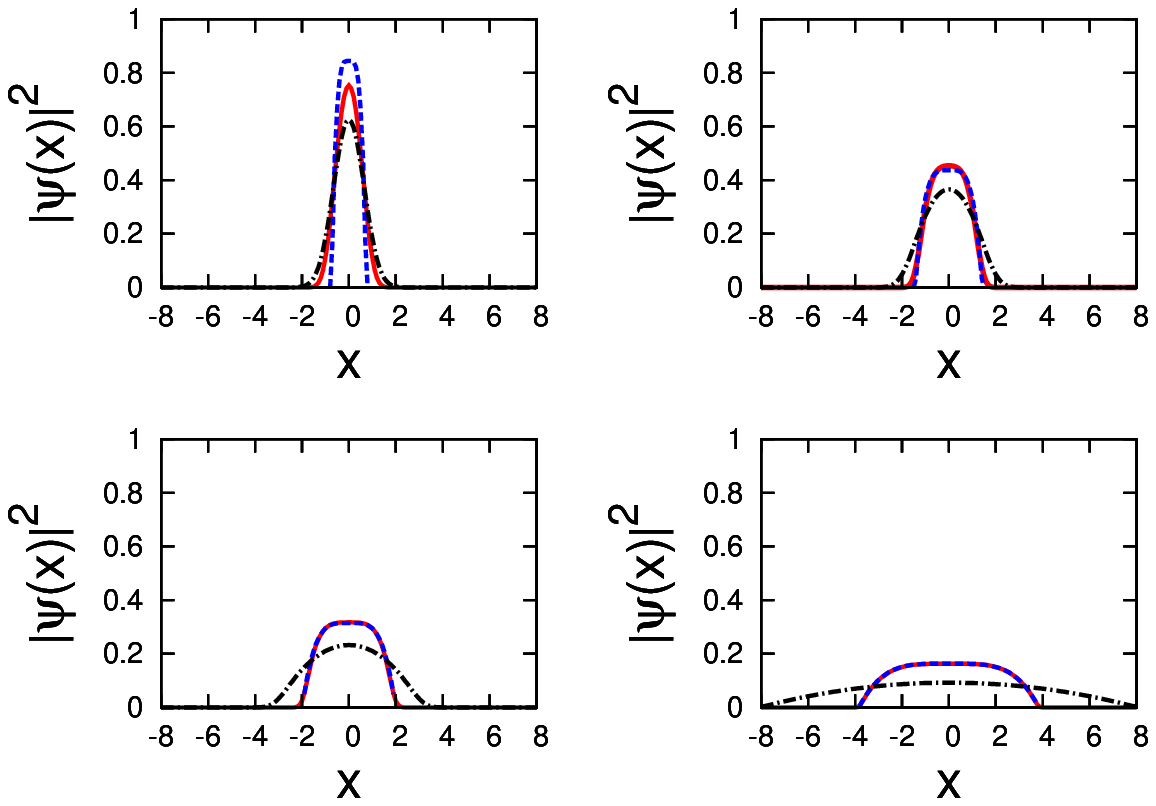}}}
\vspace*{0.5cm}
\caption[]
 {The density profiles of the condensate  ground state in 1d for
various values of N: $10$ (top left panel), $10^3$ (top right panel),
$10^4$ (bottom left panel), $10^6$ (bottom right panel). 
The results are for: the quartic trap (solid lines), TF results for the quartic
trap (dotted lines), and for the harmonic trap (dash-dot lines). 
In this and other figures x is in units of $l$.
The value of $\lambda_1$ is $1.41\times10^{-13}\,m^{2}$.}
\end{figure}
\begin{figure}
\resizebox*{4.5in}{2.5in}{\rotatebox{270}{\includegraphics{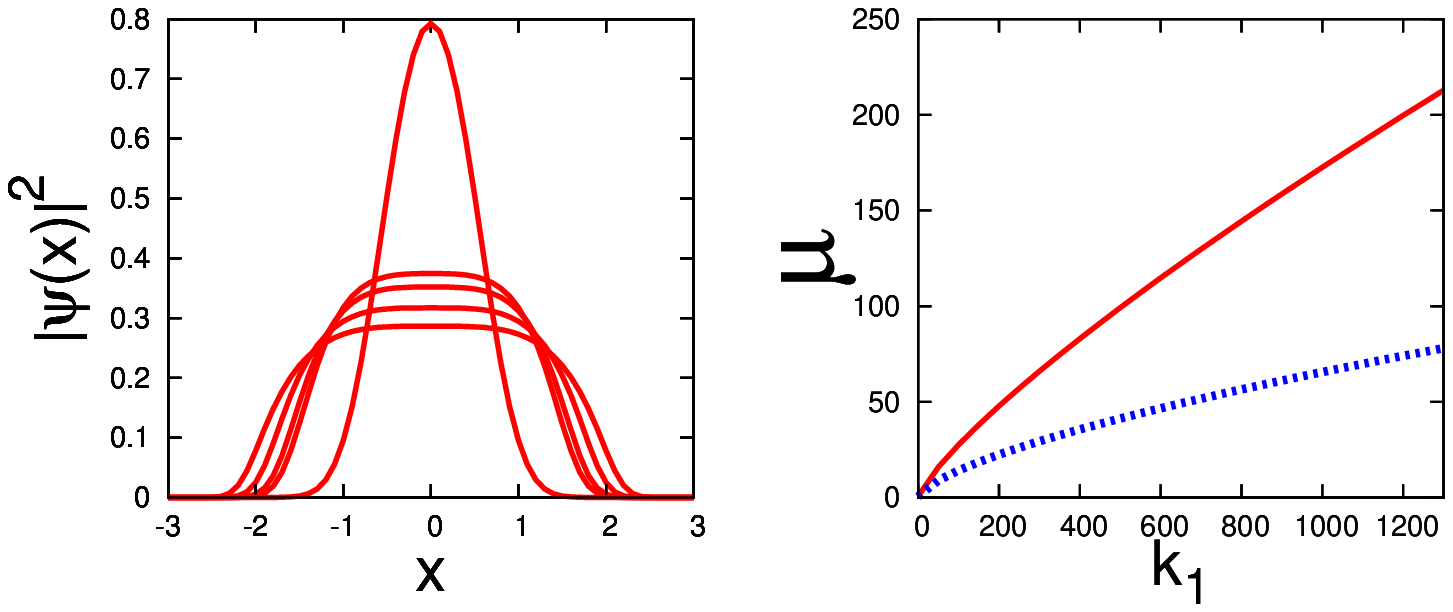}}}
\vspace*{0.5cm}
\caption[]
{Left panel: The density profiles of the condensate ground state 
of $10^{4}$ bosons in a 1d quartic potential for different values of the
interaction parameter $k_{1}$. The curves are for (from top to bottom): $k_{1}$ = $0$, $22.52$, $ 30.07$, 
$49.34$, $80.95$. Right Panel: Ground state  chemical potential $\mu$ for 
various values of interaction $k_1$ in 1d for bosons in
a quartic potential (solid line) and in a harmonic potential (dotted line). 
The $\mu$ is in units of $\hbar\omega$ and $x$ is in units of $l$.}
\end{figure}
\begin{figure}
\resizebox*{3.5in}{3.0in}{\rotatebox{270}{\includegraphics{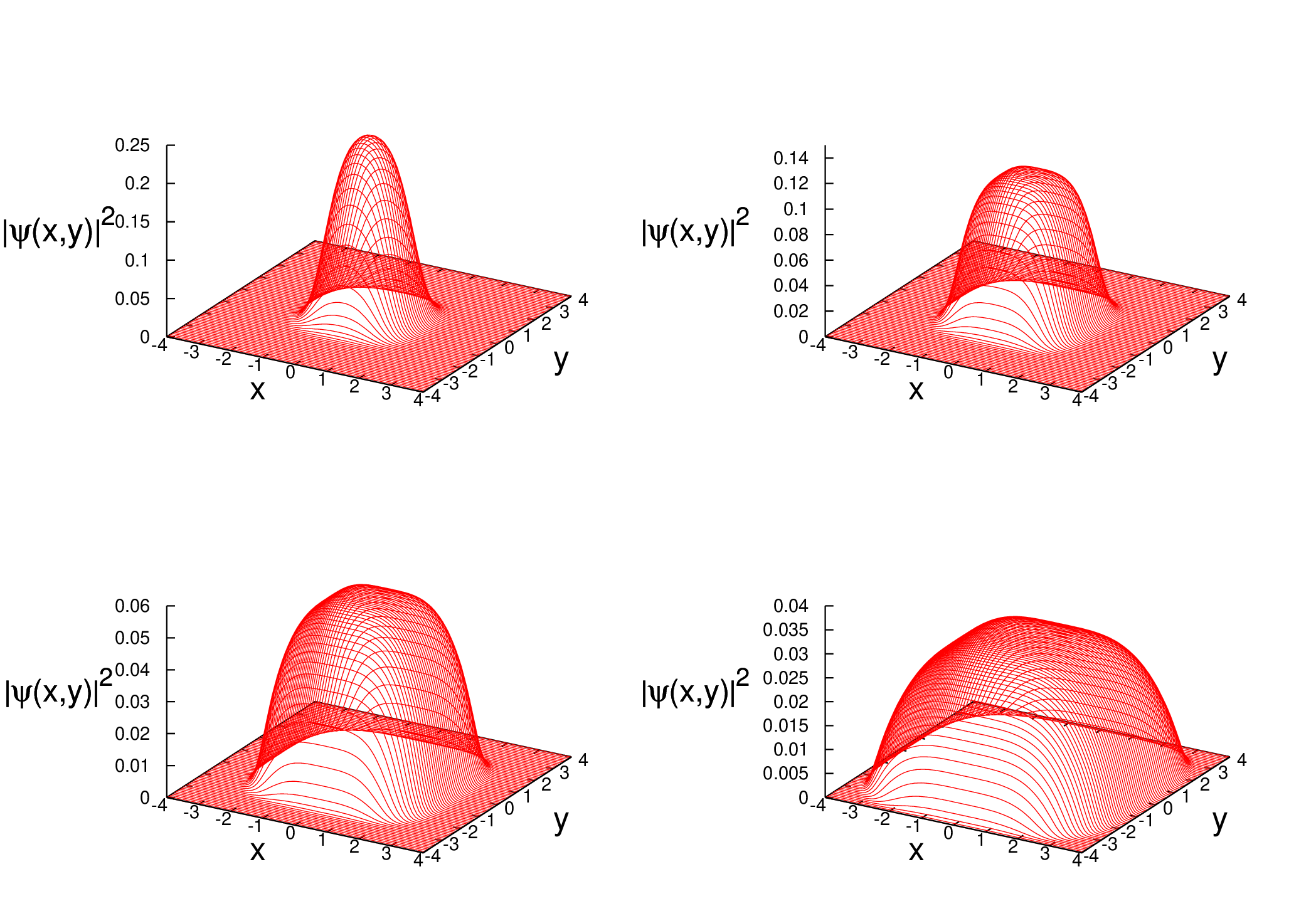}}}
\vspace*{0.5cm}
\caption[]
{The density profiles of the condensate  ground
state in 2d quartic trap for: $N=10^3$ (top left),  $N=10^4$ (top right),
$N=10^5$ (bottom left), and $N=10^6$ (bottom right). The value 
of $\lambda_{2}$ is
$6.408\times 10^{-8}\,m$. The $x$ and $y$ are in units of $l$.}
\end{figure}
\begin{figure}
\resizebox*{3.5in}{3.0in}{\rotatebox{270}{\includegraphics{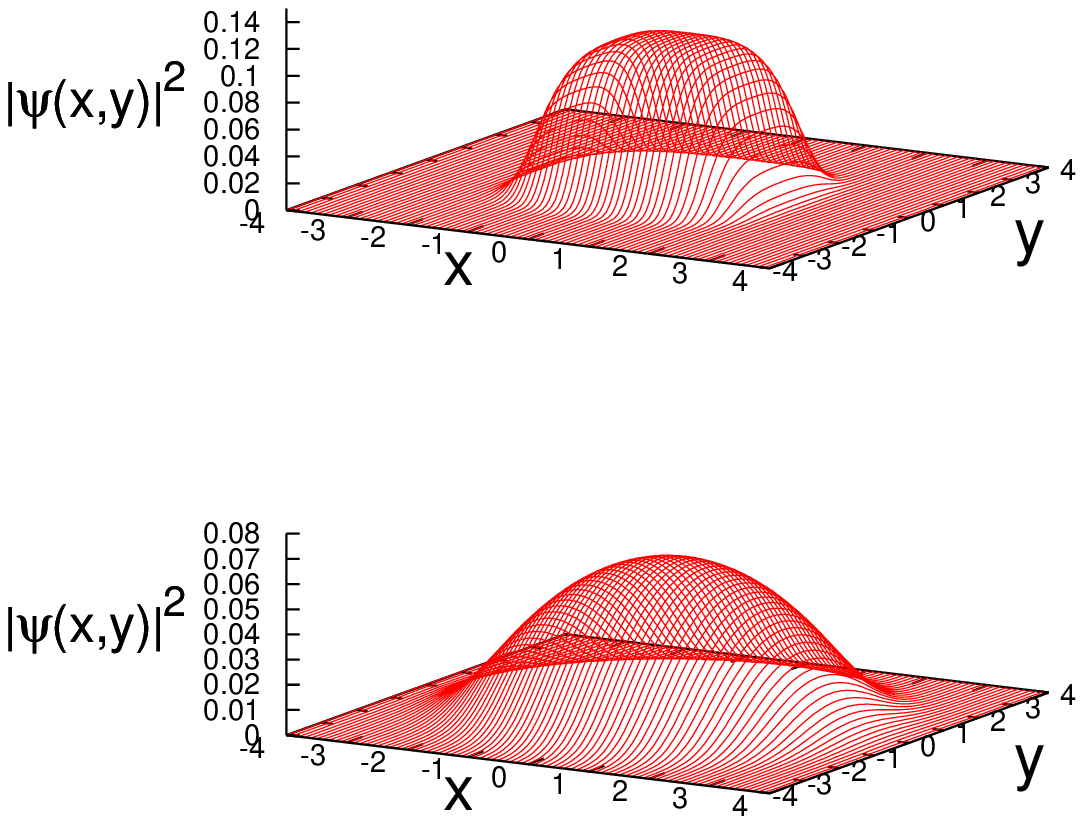}}}
\vspace*{0.5cm}
\caption[]
{The density  profiles of the condensate ground state in 2d
for N = $10^4$ for the quartic potential (top) and
for the harmonic potential (bottom). The value of $\lambda_{2}$ is
$6.408\times 10^{-8}\,m$. The $x$ and $y$ are in units of $l$.}
\end{figure}
\begin{figure}
\resizebox*{3.5in}{3.0in}{\rotatebox{270}{\includegraphics{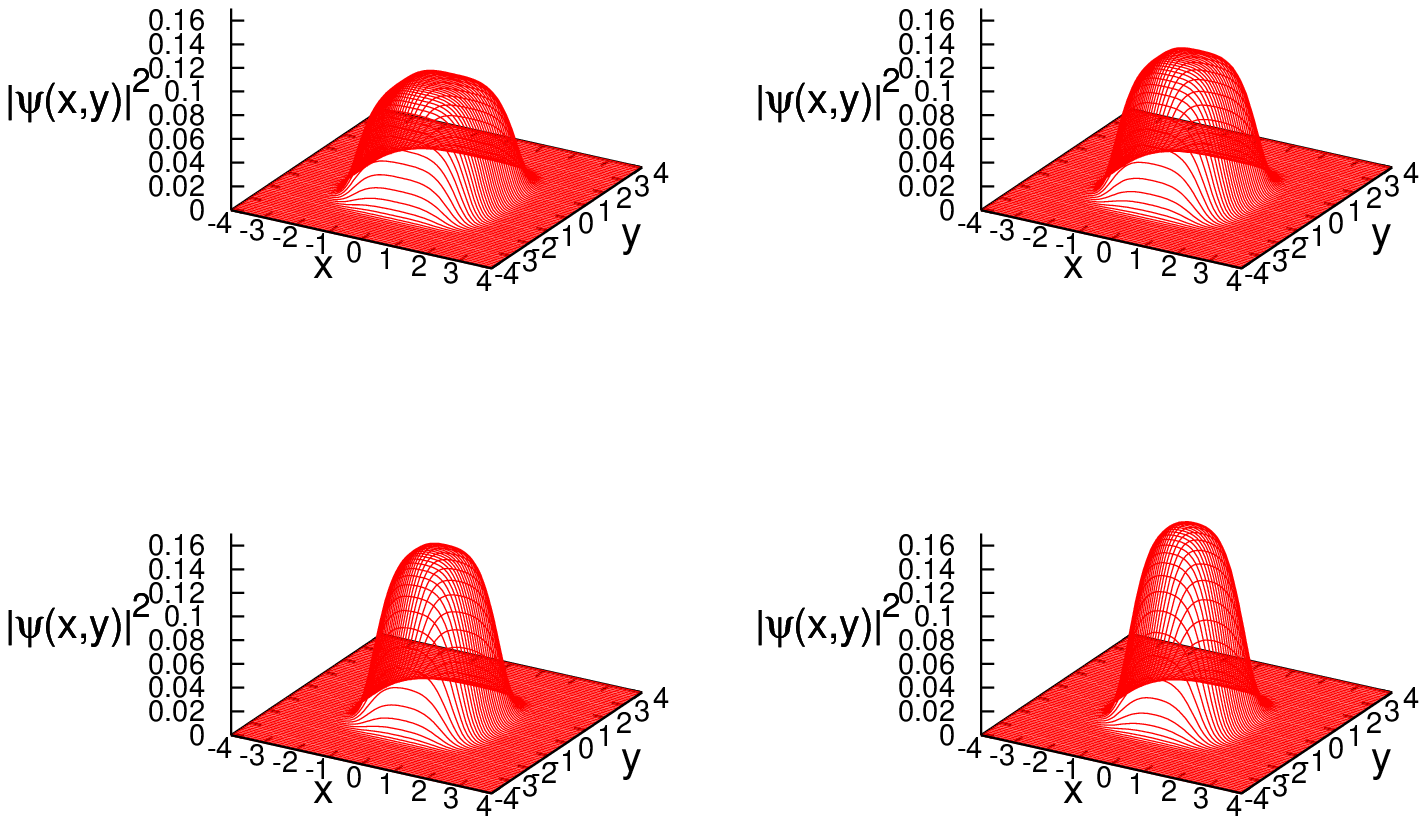}}}
\vspace*{0.5cm}
\caption[]
{ The density profiles of the condensate ground state in 2d for the quartic
potential for different values of the interaction parameter $k_{2}$.
The figures are for $N=10^{4}$ and for: $k_{2}$ = $232.25$ (top left), 
$128.21$ (top right), and $70.78$ (bottom left), $50.00$ (bottom right).
The $x$ and $y$ are in units of $l$.}
\end{figure}
\begin{figure}
\resizebox*{3.5in}{3.0in}{\rotatebox{270}{\includegraphics{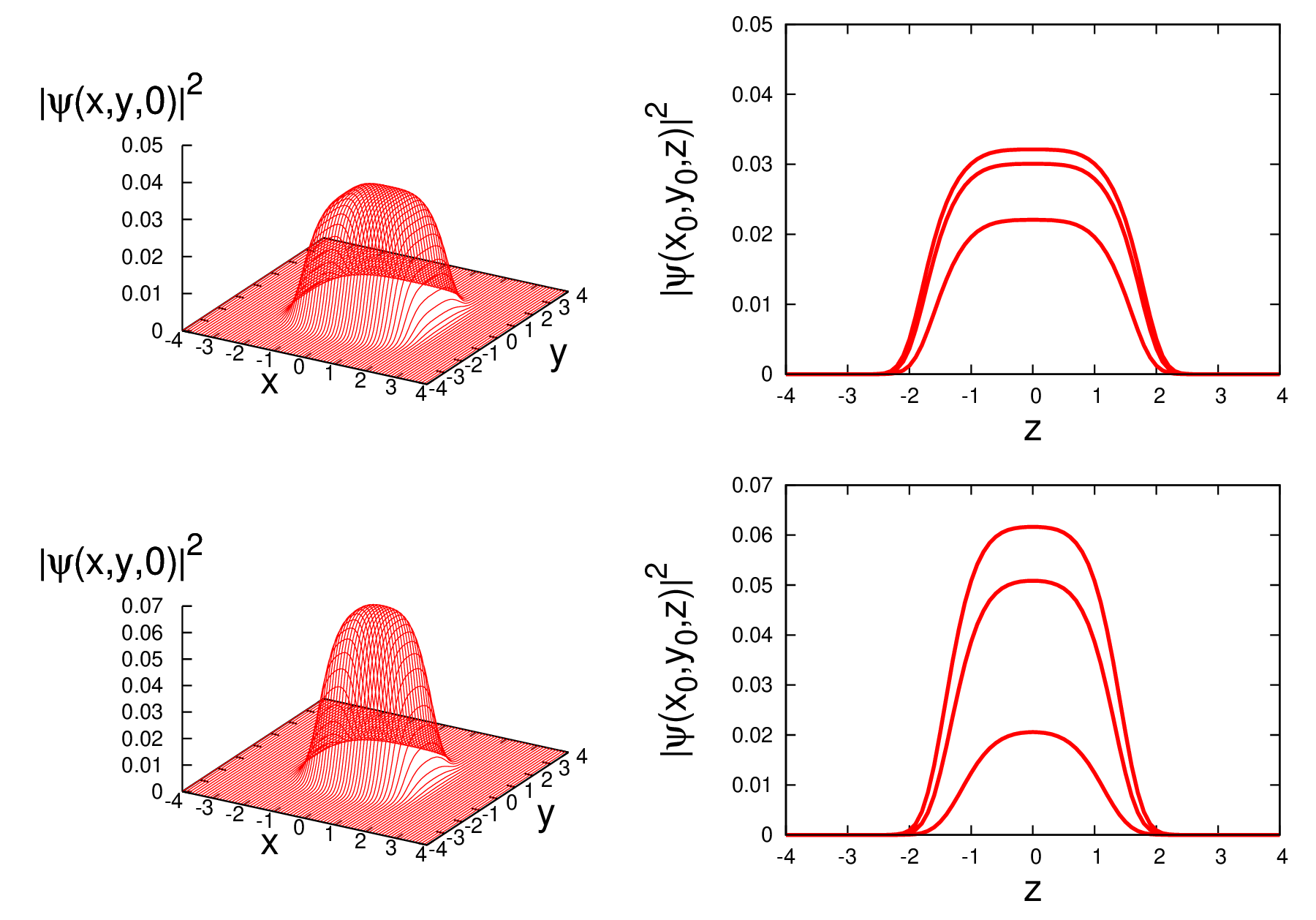}}}
\vspace*{0.5cm}
\caption[]
{ The density profiles of the condensate ground state in 3d for a quartic
potential for different values of the interaction parameter $k_{3}$.
These figures are for $N=10^{4}$ and
for: $k_{3}$ = $581.79$ (top), $145.44$ (bottom). 
 The values of ($x_{0},y_0$) are:
(0.1,0.1) (top), (0.1,1.5) (middle), and (0.1,2.0) (bottom).
The $x$, $y$, and $z$ are in units of $l$.
}
\label{scaling}
\end{figure}
\begin{figure}
\resizebox*{4.5in}{2.5in}{\rotatebox{270}{\includegraphics{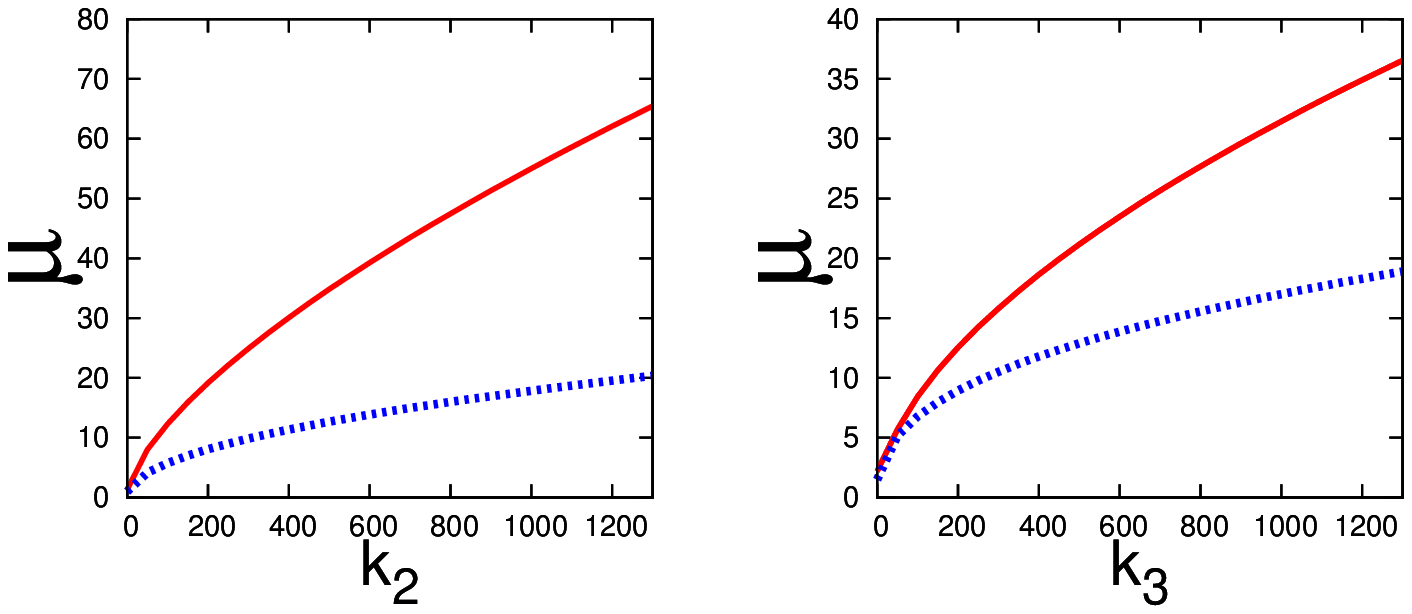}}}
\vspace*{0.5cm}
\caption[]
{A comparison of chemical potentials with interaction strengths for bosons in 
2d (left panel) and in 3d (right panel) quartic (solid lines) and harmonic 
potentials (dotted lines). The $\mu$ is in units of $\hbar\omega$.}
\label{scaling}
\end{figure}
\begin{figure}
\resizebox*{4.5in}{2.5in}{\rotatebox{270}{\includegraphics{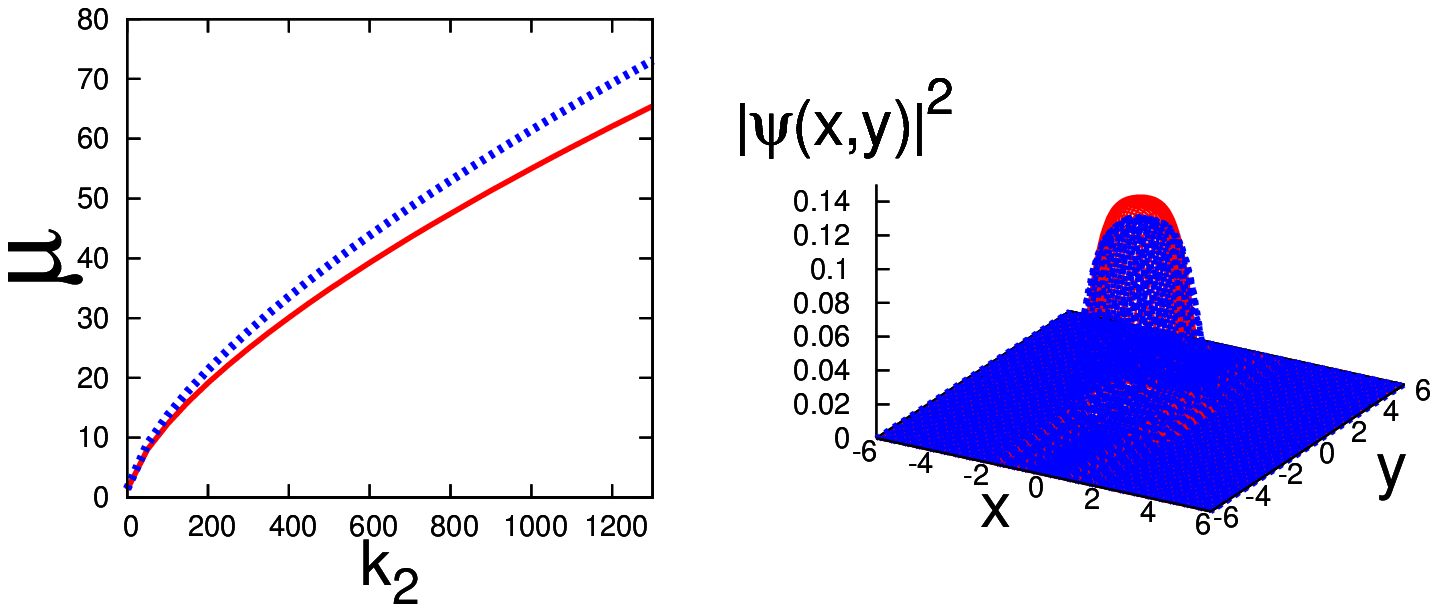}}}
\vspace*{0.5cm}
\caption[]
{Left panel: The variation of ground state chemical potential with interaction strength
$k_{2}$, in a 2d quartic potential, including cross-terms (dotted line) and 
without cross-terms (solid line). Right panel: The density profiles of 
the condensate  ground state in a 2d quartic potential for
quartic potential with cross terms (solid line) and without
cross terms (dotted line). The results shown are 
for $N=10^4$ and $\lambda_2= 6.408\times
10^{-8}m$.  The $\mu$ is in units of $\hbar\omega$ and the $x$ and $y$ are
in units of $l$.} 
\label{scaling}
\end{figure}
\begin{figure}
\resizebox*{4.5in}{2.5in}{\rotatebox{270}{\includegraphics{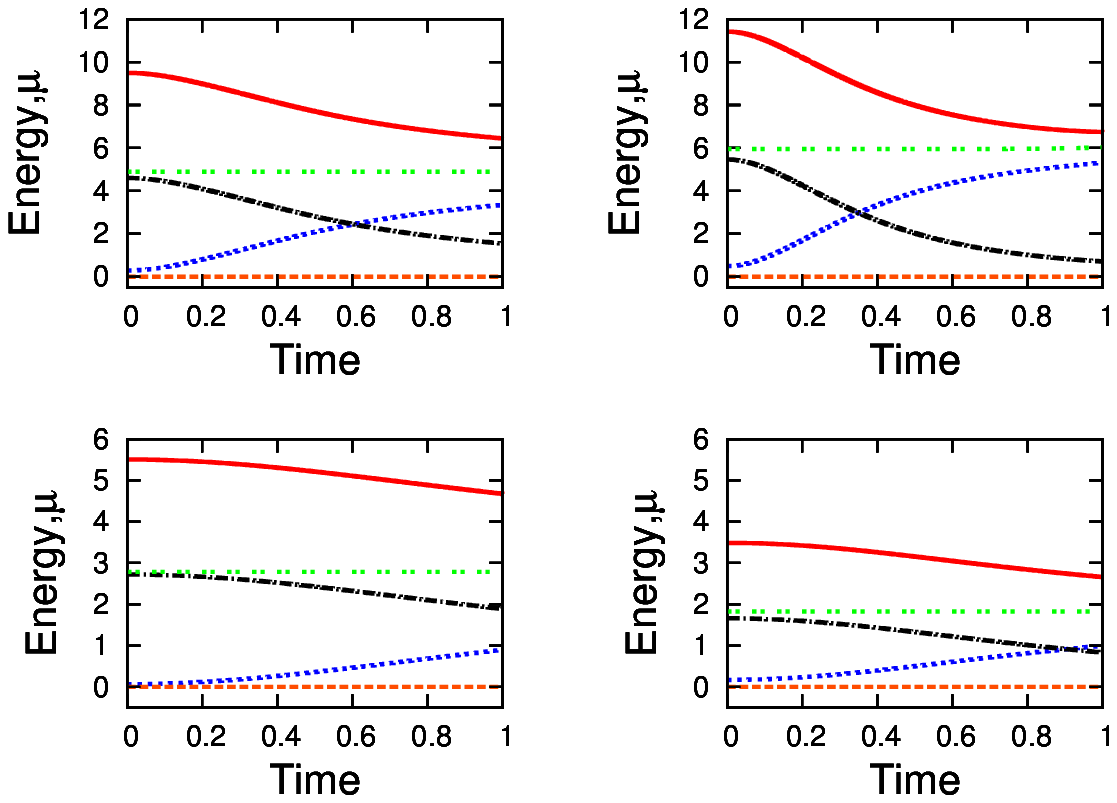}}}
\vspace*{0.5cm}
\caption[]
{The evolution of chemical potential ($\mu$) and energy components 
with time in 1d (left panels) and 2d (right panels).
The chemical potential (solid lines), the release energy (double-dot lines), 
the interaction energy (dash-dot lines),
the kinetic energy (dotted lines), the potential energy (dashed lines).
The top panels are for quartic potential and
the bottom panels are for a harmonic potential. 
The results shown are 
for $N=10^4$, $\lambda_1= 1.41\times 10^{-13}m^2$, and $\lambda_2= 6.408\times
10^{-8}m$. All the energies and $\mu$ are in units of $\hbar\omega$ and the
time is in units of $\omega$.}         
\label{scaling}
\end{figure}
\begin{figure}
\resizebox*{4.0in}{4.0in}{\rotatebox{270}{\includegraphics{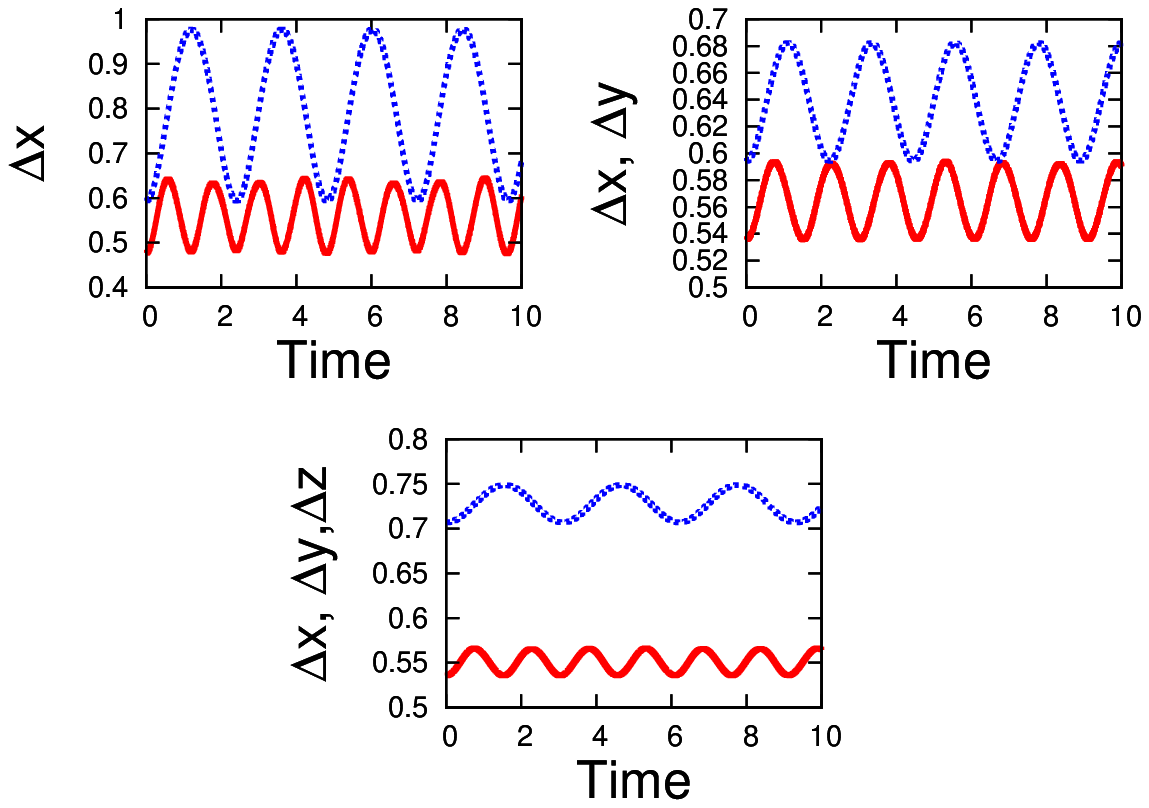}}}
\vspace*{0.5cm}
\caption[]
{
Condensate widths ($\Delta x,\Delta y,\Delta z$) as a function of 
time in: 1d (top left), 2d (top right), and 3d (bottom). The values
of $k_1$, $k_2$, and $k_{3}$  are 2.0. The solid lines are for the
quartic potential and the dotted lines for harmonic
potential. The time is in units of $\omega$.}
\end{figure}
\end{document}